\documentclass[11pt, a4paper]{article}
\pdfoutput = 1
\usepackage{graphicx}
\usepackage{amsmath}
\begin{document}
 \numberwithin{equation}{section}

\title{Thermal duality and gravitational collapse}
\author{Michael Hewitt \\
\\
Computing, Digital Forensics and Cybersecurity \\
 Canterbury Christ Church University\\  }
\date{17 February 2015}
\maketitle
\begin{abstract}

Thermal duality is a relationship between the behaviour of heterotic string models of the E(8)xE(8) or SO(32) types at inversely related temperatures, a variant of T duality in the Euclidean regime. This duality would have consequences for the nature of the Hagedorn transition in these string models. We propose that the vacuum admits a family of deformations in situations where there are closed surfaces of constant area but high radial acceleration (a ‘string regularized’ version of a Penrose trapped surface), such as would be formed in situations of extreme gravitational collapse. This would allow a radical resolution of the ‘firewall paradox’ by allowing quantum effects to significantly modify the spacetime geometry around a collapsed object. A ‘string bremsstrahlung’ process would convert the kinetic energy of infalling matter in extreme gravitational collapse to form a region of the deformed vacuum, which would be equivalent to forming a high temperature string phase. A heuristic criterion for the conversion process is presented, relating Newtonian gravity to the string tension, suggesting an upper limit to the strength of the gravitational interaction. This conversion process might have observable consequences for charged particles falling into a rotating collapsed object by producing high energy particles via a variant of the Penrose process. 
 \\


\end{abstract}
\
\

\section{ Black holes and discrete symmetries}

The family of classical black hole solutions is parameterised by mass, charge (standard model gauge charges) and angular momentum. Consider the effect of discrete symmetries on them. Whereas C and P are satisfactory T is not, as time reversed black holes are not (usually) allowed. This represents a radical violation of T and by extension CPT symmetry, and because of this the antiparticles of black holes are problematic to define. Black holes (BH) can in principle be produced in scattering experiments. The highest energy cosmic rays have about enough energy ($\sim 10^{19}$ GeV) to form quasi- stable black holes on collision. Short lived black holes could be produce at the LHC in some scenarios \cite{LHC} so that time reversal symmetry is not an entirely theoretical question. The resolution which we propose to the problem of realising time reversal symmetry exploits thermal duality, a discrete symmetry of certain string models. 

\section{Time reversal and the black hole interior}
The exterior solution for a non-rotating BH is T symmetric, e.g. in the neutral case

\begin{equation}
ds^2 = (1-\frac{R_s}{R})dt^2 - (1-\frac{R_s}{R})^{-1}dR^2 - R^2(d\theta^2 + \sin^2\theta d\phi^2) 
\end{equation}

In the rotating case, T is equivalent to P in the exterior region, and so still produces a permitted state. An inward continuation is possible, and indeed would be necessary to roll the solution forward in time, but the resulting e.g. Eddington-Finkelstein metrics \cite{Eddington}, \cite{Finkelstein} necessarily violate T symmetry. Time becomes correlated with topology, in that events inside the horizon can be later than those outside, but not vice-versa. 

The quantum version of a black hole (a Hawking BH) can be in thermal equilibrium with an external heat bath in what appears to be a T reversible condition to an external observer. This suggests that the final state for gravitational collapse should be (statistically) T reversible.

\section{Black holes as particle accelerators}
Black holes can be seen as being the ultimate particle accelerators. Collapsing matter acquires a divergent centre of mass energy, and this is also the case for individual particle pairs. This is related to the necessity of a singularity under generic conditions as shown in the Penrose singularity theorem. Particles are isolated by encountering a space-like singularity. We may say that time must end abruptly at a finite world-line time in order to prevent infinite energy collisions. Note that the singularity theorem indicates that a problematic situation is set up as soon as a trapped horizon is formed and crossed.\

If the energy released by the gravitational accelerator were converted to string, the possible number of states produced would increase exponentially with energy, due to the form of the string spectrum. Within a certain (fixed) distance of the horizon, the effective string entropy would match and then exceed the Bekenstein-Hawking entropy \cite{Hewitt93} \cite{Susskind93} .

\section{Black hole information problems}
Information carried by the collapsing matter is destroyed at the singularity. New information is randomly created  during the Hawking process, and the entangled pair particles which fall into the hole are also destroyed at the singularity. Reconciliation between the information going in and coming out is difficult to achieve because of the monogamy of entanglement in quantum information theory \cite{Almheiri} \cite{Braunstein}.\

\begin{figure}
\begin{center}
\includegraphics [height = 120mm, width = 100mm]{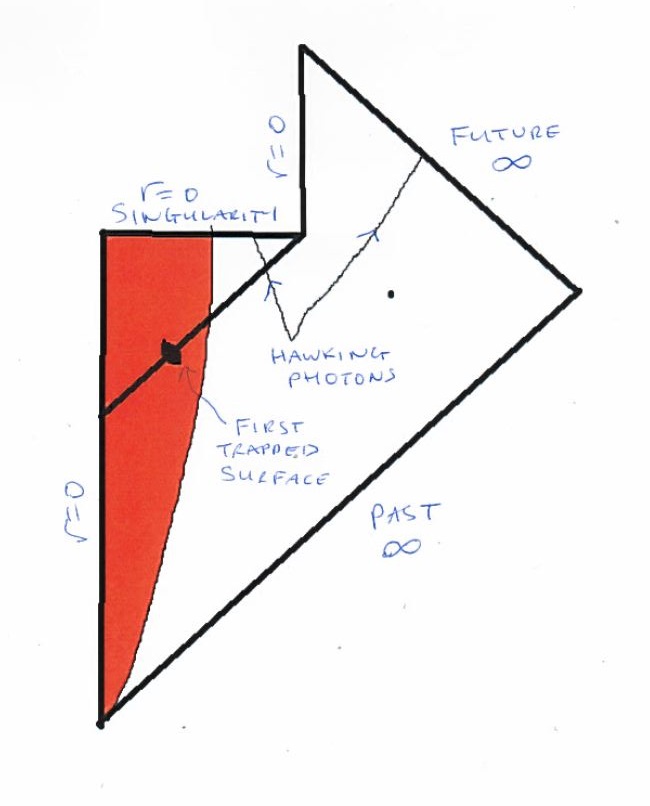}
\end{center}
\caption {Penrose diagram for an evaporating black hole}
\end{figure}
Note the causal disconnection between the original information and the late Hawking radiation as both are terminated at the singularity. 
Both creation and destruction effects are inconsistent with quantum information theory, and both are related to the T asymmetry of the BH. Common sense suggests some kind of re-emergence of the original information, but studies show that this is difficult to achieve – this is the ‘firewall paradox’  \cite{Almheiri}, \cite{Braunstein}. Some kind of near horizon buffer apparently needed to store the information.\

The relationship between the Bekenstein-Hawking entropy and the horizon entropy is unclear. Whereas the Bekenstein-Hawking entropy is finite and depends only on the area  $S=A/4$, the thermal entanglement entropy cut off at distance $r$ from the horizon
\begin{equation}
S = \frac{NR^2}{90r^2}
\end{equation} 
diverges as $r \rightarrow 0$, and depends on the number of particle species. 

If we accept the holographic principle \cite{t'Hooft93}, \cite{Susskind94}, then even immediately after crossing the horizon the information density (by surface area) of the collapsing region exceeds the holographic bound based on the Bekenstein-Hawking entropy. Before reaching the singularity, the conventional entropy of the collapsing matter will likewise exceed the Bekenstein-Hawking bound for an enclosing surface. Any holographic bound must in any case be violated in the approach to a singularity with zero surface area. A positive result from the Fermilab Holometer experiment \cite{Hogan} would raise the status of this issue as further evidence of a conflict between the conventional scenario and quantum information theory. Taken together with analysis of the firewall problem, this suggests that a new physical mechanism is needed to store information just outside where a horizon would form during collapse. More specifically we propose that such a mechanism could operate just outside where a trapping horizon (in the sense of Penrose \cite{HawkingEllis}) would form. While non-conservation of quantum information and infinite energy particles suggest that a true black hole may be physically impossible, it seems that the near-horizon neighbourhood is unremarkable unless some novel effect intervenes.

\section{Black holes: A stage illusion?}

Perhaps the black hole is like a cosmic stage illusion, in that we are in a sense misdirected by reasonable assumptions to believe that something impossible has happened. As in many such illusions, the key is that we miss the possibility that the trick has already happened earlier than we think. In this case, that the infalling matter is converted to another, radically different form before a closed trapping horizon can form.
In the following sections, we will take thermal duality seriously and look at the consequences, the equivalence principle is valid. However, we do not assume that the space-time near collapsed object is necessarily ‘normal’. Key features of the final state in our model are that it is embedded in and in equilibrium with a surrounding vacuum equivalent to that of a conventional black hole. The thermodynamic properties of Hawking radiation can then be used as a guide. The final state is a statistically time reversible thermal (ensemble) close to equilibrium with its surroundings, and which will slowly evaporate if surrounded by a vacuum.\

We conjecture that it is possible to excise the near horizon region of a black hole and replace this with a hot black body. Proposal is that regions of a hot string phase can replace black holes. Further, we will present a mechanism that could convert collapsing matter to this hot string phase, averting the production of BHs.

\section{The Hagedorn transition}
The nature of the Hagedorn transition \cite{Hagedorn} is that strings of arbitrary length form. The transition takes place at the temperature where the thermalon or winding mode in Euclidean time becomes massless \cite{AtickWitten}. Condensate formation makes sense of Hagedorn point as a phase transition, and enables a continuation to higher temperatures to be made. The effective Lagrangian for the hot phase for small deformations $\phi $ is \cite{AtickWitten}
\begin{equation}
\mathcal{L} = \frac{1}{2}\sqrt{-g}\partial_{\mu} \phi \partial^{\mu} \phi + \frac{1}{2}m^2(\beta)\phi^2 + \frac{1}{4!}\lambda (\beta) \phi^4 + O(\phi ^6)
\end{equation}

A distinctive feature of heterotic models is thermal duality, or $T$ duality in Euclidean time compactification \cite{AtickWitten}, \cite{Sathiapalan}, \cite{ObrienTan} . Such models do not have D branes \cite{Polchinski} so the model is purely string based.\

Thermal duality is shown by the characteristic behaviour of thermalon trajectories in left-right moving energy as temperature increases, see Figures 2 and 3.
\begin{figure}
\begin{center}
\includegraphics [width = 100mm]{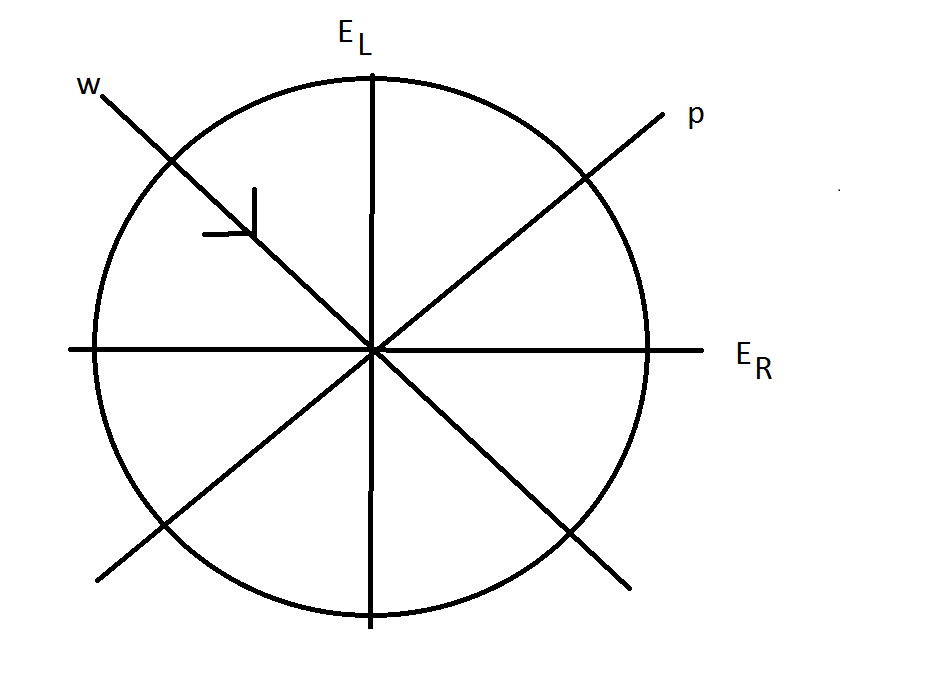}
\caption{Non-heterotic tachyon trajectory}
\end{center}
\end{figure}

\begin{figure}
\begin{center}
\includegraphics[width = 100mm]{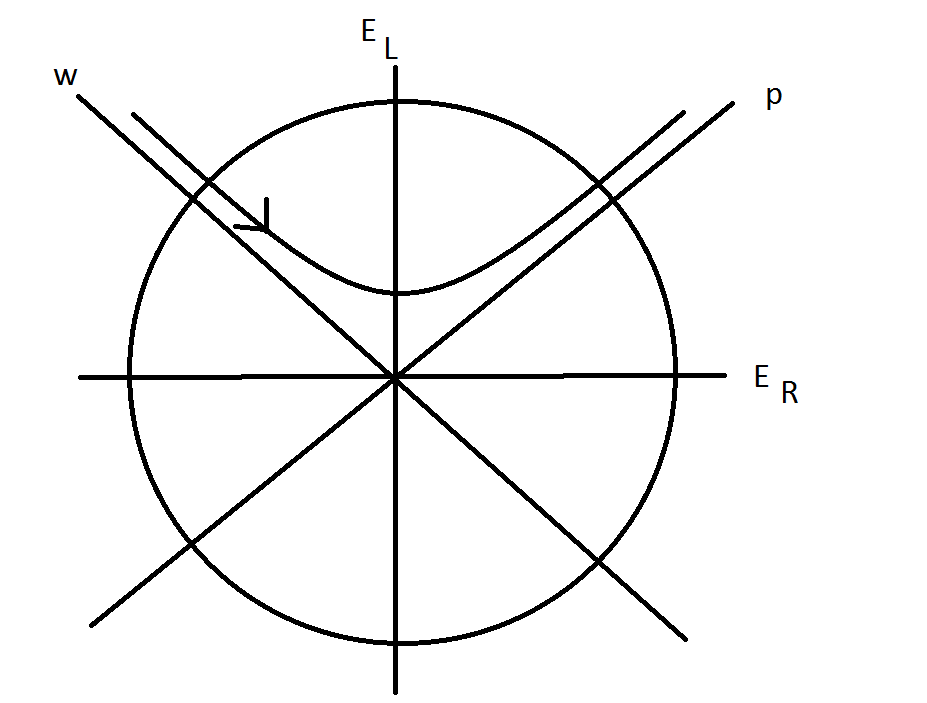}
\caption{Heterotic tachyon trajectory}
\end{center}
\end{figure} 

Thermalons are expected to interact with normal matter, as they regulate the thermal string spectrum by increasing the effective string tension \cite{Hewitt02}. However, the thermalon mass diverges as $T$ approaches 0, so there are no long range forces between ordinary particles associated with them.\

The amplitude is on-shell at the upper and lower transition points, but there is an issue off-shell, in that the scattering amplitude is only relevant on shell and so a full calculation is  out of range of these simple method. Nevertheless, we will use only qualitative features of the couplings, so the analysis should hold.\

To describe the final state, the bulk may be approximated by a space filling solution \cite{Hewitt2014}, which gives an almost constant temperature in the bulk. Expulsion of gravitational gradients is a consequence of the thermal duality of $Z$. The surface energy is positive on the outside, and negative on the (unstable) inside of a shell region. The bulk has a hyperbolic geometry:
 \begin{equation}
du^{2} =  \frac{dr^{2}}{1+a^{-2}r^{2}} + r^{2}(d \theta^{2} +
\sin^{2} \theta d \phi^{2}) \label{BL}
\end{equation}

so that the bulk properties are area dependent (or specifically, area difference dependent). The final state is in equilibrium (with surrounding normal vacuum) and has T symmetry. The final state has holographic properties, as information is stored in a quasi 2-dimensional form (only a few hundred Planck lengths thick). There may be a connection to non-commutative geometry \cite{t'Hooft} in particular the fuzzy sphere \cite{Madore}. The shell may be modelled a quotient of fuzzy spheres \cite{Hewitt2014}.\

The free energy of a BH and a broken symmetry string region should be the same (to first order in $1/M$) as Hawking’s entropy calculation applies to both, i.e. half of the total energy in the Schwarzschild case. By comparison a Hagedorn string has zero free energy. This model has positive free energy in the surface layer, making agreement possible.

\section{Thermalon weak field solutions} 
These are relevant to the beginning of the conversion process, and show that there is no energy barrier to the initiation of the conversion process. 
The thermalon field equation in Rindler coordinates \cite{BirrellDavies} is:
\begin{equation}
\frac{\partial^2 \phi}{\partial r^2} + \frac{1}{r}\frac{\partial \phi}{\partial r} = m^2\beta (r) \phi = (\frac{1}{4r^2}+ 4r^2 -6)\phi \label{DE}
\end{equation}
where the thermalon mass $m$ is given as a function of $r$ by
\begin{equation}
m^2(r) = (\frac{1}{4r^2} +4r^2 -6)
\end{equation}
This admits the following finite accelerating wall solution
\begin{equation}
\phi =\epsilon \sqrt{r} \exp (-r^2) \label{sol}
\end{equation}
where $\epsilon$ is a small parameter.

\begin{figure}
\begin{center}
\includegraphics [height = 50mm, width = 100mm]{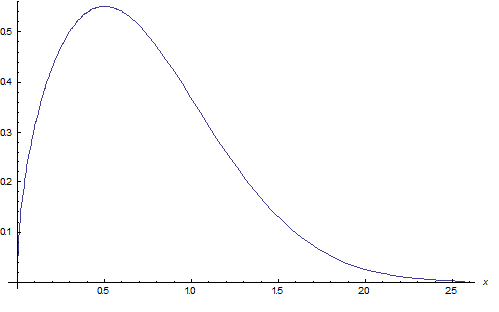}
\end{center}
\caption {$\phi / \epsilon$ for the weak heterotic solution}
\end{figure}
We can calculate the back reaction on the metric as follows. The gravitational source  $\mu = \rho +\ Tr( p)$  is given by
\begin{equation}
\mu = \frac{1}{2}\frac{\partial ( \theta V)}{\partial \theta}
\end{equation}
where $\theta = \beta^{-2} = T^2$. From this we can calculate the red-shifted gravitational source, as seen from infinity.
Introducing a factor of $1/\beta_\infty= 8 \pi M$ to compensate for the redshift seen by an observer at infinity gives

\begin{equation}
\frac{\beta \mu}{\beta_\infty} = 8\pi^2 M e^{-2r^2}(\frac{1}{4}-3r^2)\epsilon^2/\beta_0
\end{equation}
and the total source as seen from infinity becomes

\begin{equation}
\frac{1}{\beta_\infty}\int_{0}^{\infty}\beta \mu dr = 4\pi M (\frac{\pi}{2})^{3/2}\epsilon^2
\end{equation}
which is finite and positive.

\begin{figure}
\begin{center}
\includegraphics [height = 50mm, width = 100mm]{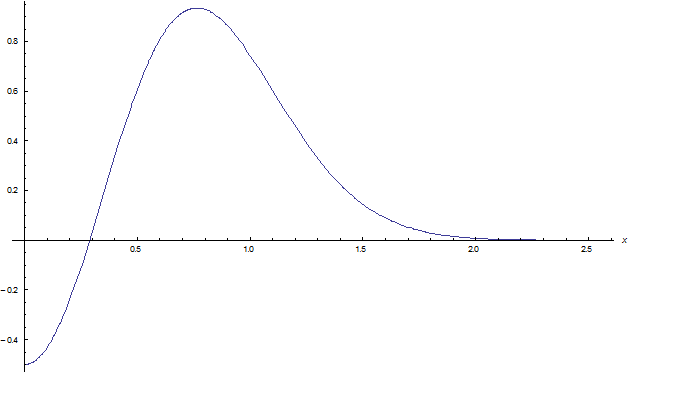}
\end{center}
\caption {Red shifted gravitational source}
\end{figure}

Note the negative energy density close to the horizon. The source $\mu$ is zero at $ r = \frac{1}{2\sqrt{3}}$ and is negative to the high temperature side of this point and positive to the low temperature side (see Figure 5). The wall has a warped geometry with area difference across the wall given by

\begin{equation}
2\pi \delta A \sim 4\pi GA \int_{0}^{\infty}\mu \beta dr = 2\pi GA(\frac{\pi}{2})^{3/2}\epsilon^2
\end{equation}
This can be described by the warp factor $w$ where

\begin{equation}
w-1 = G (\frac{\pi}{2})^{3/2}\epsilon^2
\end{equation}
The mechanics of the solution are unusual. A high pressure region is squeezed between accelerating boundaries with positive (leading side) and negative (trailing side) inertia. Newtonian mechanics allows this configuration to undergo self-sustained acceleration. However, stable thermalon distortions are impossible against a flat background due to the warp factor. The wall does not fit, and a section will be unstable (the geometry will necessarily change during propagation). This can be seen in another way from the requirements of energy and momentum conservation. However they become possible near to a collapsed object. This geometric feature allows a conversion mechanism from matter to deformed regions, but only near collapsed objects. Accelerated thermalon modes will remain unexcited in empty space.

\section{Thermalon string traps}
There can be stable thermalon string traps near collapsed objects. Temperature defined relative to local timelike Killing vector. Deformations give interpolation between exterior Schwarzschild solutions with different masses. The thermalon string trap around a black hole could be excited in a non-zero deformation to produce a kind of enhanced black hole. This would give an increased surface area and mass to the collapsed entity. The number of possible quantum states would increase, so that the hole would acquire ‘stringy hair’ and the entropy would increase in proportion to the area difference across the thermalon region, in a way that would maintain consistency with Hawking’s  area law.\

So far, this gives a possibility of thermalon energy suspended near a horizon, giving a dressed black hole which requires additional information to specify its quantum state. There would still, however, be a black hole at the centre, with nothing to prevent particles from falling in.  The ‘weak field’ thermalon shell solution can be extended to make it possible to eliminate the black hole, and indeed to circumvent the possibility of their forming during gravitational collapse.
\subsection{Space filling solution}
A solution to the thermalon field equation is possible at the values $\beta_s, \phi_s$ for $\beta, \phi$ where $T^2V$ is a minimum so that the gravitational source $\mu$ vanishes, as such a point must exist by the mean value theorem, giinge a stable space-filling solution with a hyperbolic spatial geometry

 \begin{equation}
du^{2} =  \frac{dr^{2}}{1+a^{-2}r^{2}} + r^{2}(d \theta^{2} +
\sin^{2} \theta d \phi^{2}) \label{BL}
\end{equation} 
\subsection{Ball solution}
A finite region with interior approximately as above. Such a solution features expelled gravitational gradients. The London response \begin{equation}
\frac{\partial \mu}{\partial \beta} 
\end{equation}  \cite{Hewitt2014} has opposite signs on either side of the stable temperature $\beta_s$ in the bulk region. This has the effect of expelling gravitational gradients from the bulk, so that the gravitational source is confined close to the boundaries and the high temperature phase becomes effectively a kind of gravitational (magnetic) superconductor.
\subsection{Intermediate cases}
The transition to strong $\phi$ gives increasing nonlinearity with increasing area difference and thickness of the shell. This is analogous to the situation of a pendulum which becomes increasingly nonlinear as the amplitude increases, and the solutions (elliptic functions) become increasingly flat at the extremes. The shells can be parameterised conveniently by the warp factor, which may vary across the surface for collapses which are not spherically symmetric. The opacity of the shell will increase as it thickens, approaching the condition of a black body.
For a realistic conversion process, we must meet the following requirements:
1. Locality of the conversion process, and 2. Conversion of kinetic energy is possible for a single infalling particle.\

\section{Thermalisation} 
We propose a  'space crushing' model of the transition mechanism. Infalling matter couples to thermalon string traps where trap formation is due to the combined gravity of the infalling matter, so that trap formation is a cooperative phenomenon. On arriving at a certain distance (estimated below) from a potential trapping horizon, the particle will produce a new critical acceleration stable surface, effectively focussing the local thermalon traps into a kind of standing wave. The trap then crushes space between this outer surface and the current location of the particle, as the kinetic energy of the particle is converted and increases the warp factor of the trap. The trap then becomes analogous to a crumple zone which absorbs the energy released during gravitational collapse, resulting in thermalisation. Matter coupled to a trap will appear to be almost static to an outside observer, but moving close to $c$ to an interior one. \

\section{Deformation generation mechanism}
We will look next at the possibility of a self-consistent mechanism to feed energy from decelerating matter to the thermalon field on formation of a stable trap. The 4 puncture worldsheet diagram gives the simplest matter-thermalon interaction. Making a comparison to the Larmor radiation formula  $P \sim (\dot{J})^2$ , note that the thermalon couples to the  left moving component of energy on the string worldsheet for heterotic strings. At a trap, $E$ and $m$  are $\sim \sqrt{Mm}$ and $\sim G\sqrt{Mm}$ respectively, so that $P \sim 1/G$ here, and this is consistent with $P \sim GP^2$, the analogue of the Larmor formula. It is therefore apparently possible for power to drain from a decelerating particle into a static exterior thermalon trap, and when possible, such a process will be thermodynamically favourable (and so overwhelmingly likely) because of the large release of entropy. 
Thus spontaneous deceleration gives reinforcing particle trajectories, in addition to the conventional free-fall type in the Feynman picture of quantum processes, with decelerating trajectories dominant because of the higher overall probability from the large number of final states (the reduction in free energy during conversion  shows that this is a dissipative process).

\section{Newtonian formalism}
Where will a trap with a static exterior form during gravitational collapse? Consider the following simple cases, which will indicate a simple approximate criteron for the nucleation process to be initiated.

\subsection{Spherical collapse}
Consider a thin spherical shell collapsing inwards, which will be partly realistic due to extreme Lorentz contraction relative to a static observer. A spherical shell of mass $m$ falling onto an existing collapsed object of mass $M$ will cause a static trap to form at a distance $d \sim \sqrt{Mm}$ from the surface, due to the form of the Schwarzschild neck geometry. The solution peels from the outside, with the mass of the shell remaining critical compared to its area as it contracts. Kinetic energy is progressively converted into warp form as the collapse proceeds, and space becomes longitudinally compressed to form a hyperbolic spatial geometry.\

\subsection{Single infalling particle}
A point particle of mass $m$ falling onto an existing collapsed object of mass $M$ will increase in energy relative to local static observers, so that at a distance $d$ from the surface it has energy 
\begin{equation}
E = \frac{GMm}{d}
\end{equation}
 This will form a horizon which will meet $M$ when 
\begin{equation}
d \sim G \frac{GMm}{d}
\end{equation}
 and again $d \sim G \sqrt{Mm}$. 
The critical gravity surface of the massive body is tidally distorted and develops a ‘finger’ towards the infalling particle and nucleation begins as this finger and the particle meet. The geometry is similar to numerical simulations of merging black holes. The trap surface will evolve towards a spherical form in the case of zero angular momentum. Energy will diffuse around the trap, with the radial compression initially greatest near the particle location.\

In this case, the increment in area $2Mm$ of the trap is comparable to the area over which thermalons can spread during the conversion time $t \sim \sqrt(Mm)$, again confirming the significance of this scale.\

\subsection{Colliding particles}
In the case of very high energy colliding particles, both will become nucleation sites when their mutual gravity becomes $\sim T$. The subsequent evolution will be similar to the previous case.

Re-arranging the formula for stable trap formation in all of these cases gives 
\begin{equation}
F \sim \frac{GMm}{d^2}
\end{equation}
On restoring the string tension scale, the criterion for stable trap formation becomes $F \sim T$  i.e. the Newtonian gravitational force becomes comparable to the string tension as the two bodies approach. This shows the mutual nature of the trap formation process, and that it is a kind of epiphenomenon of the gravitational interaction. It also addresses one of the most egregious problems of ‘firewall’ scenarios, in that the infalling body must apparently encounter something in empty space to interact with.  Here the formation of a stable trap which initiates the phase transition is caused by the joint effect of the two bodies, so that the presence of the infalling body is needed for the conversion conditions to be achieved.

\section{Energy time and distance scales}

Applying the criteria above to the specific case of an electron falling nto a 10 solar mass object gives an energy release of $10^{26} J$ and a distance scale for nucleation of $10^{-26} m$.\

The time taken to conversion to a hyperbolic region is $O(M^2)$, and the time for diffusion to complete across the surface for a non-spherical collapse is similar whereas the time for evaporation by the Hawking process is $O(M^3)$. Shell conversion would still be underway for collapsing astrophysical objects in the current epoch. The thickness of the shells would be $O(M)$, or only a few hundred Planck lengths, making them effectively 2 dimensional 'holographic' objects. 

\section{Rotating case}
The deceleration of charged particles (braking effect) should produce photons, and in the rotating case their energy will depend on direction. Pairs produced in forward and backward directions may give a Penrose energy extraction mechanism. The outgoing photon energy (measured from infinity) is
\begin{equation}
E \sim p \frac{a}{2M}
\end{equation}
for small a (angular momentum per unit mass) or
\begin{equation}
E \sim p
\end{equation}
 in the near critical case. There may be a possibility of producing high energy cosmic rays, if these can be scattered outwards.

\section{Conclusions}
There is no horizon and so no divergent entanglement entropy in these models. The field (for level 0 strings) state however becomes entangled with the long string sectors, and so the fields will carry incomplete state information. As no singularity forms, the problems of divergent particle energy and discontinuity of quantum information are avoided in this scenario. Further investigation is needed to determine whether high energy particles can be produced in the rotating case.

\end{document}